# Uniform maximum projection designs for computer experiments

Miroslav Vořechovský*, Jan Mašek

*Institute of Structural Mechanics, Brno University of Technology, Veveří 331/95, 602 00 Brno, Czech Republic*

## HIGHLIGHTS

- Introduce periodic-distance MaxPro (uMaxPro) that ensures statistical uniformity.
- Remove edge effects while preserving MaxPro's projection-aware space-filling.
- Deliver unbiased integration and lower RMSE with small simulation budgets.
- Improve uniformity in all subspaces; outperform Halton in tested benchmarks.
- Provide open-source code and design libraries for engineering adoption.

## ARTICLE INFO

## ABSTRACT

Space-filling experimental designs are widely used in engineering computer experiments, where only a limited number of expensive model evaluations can be afforded. Distance-based designs such as Maximin or Minimax ensure global space-filling, while Latin hypercube sampling enforces uniform one-dimensional projections, yet neither guarantees uniformity in low-dimensional subspaces. Maximum Projection (MaxPro) designs were introduced to improve uniformity in low-dimensional subspaces, yet their original formulation relies on the Euclidean distance and may induce systematic density distortions in bounded domains. We demonstrate that the standard MaxPro criterion leads to statistically non-uniform sampling, resulting in undersampling of corner regions and biased Monte Carlo estimates.

To remedy this issue, we introduce a periodic variant of the criterion, termed Uniform Maximum Projection (uMaxPro), in which the Euclidean metric is replaced by a periodic distance based on the minimum image convention. The proposed uMaxPro designs preserve the projection-aware structure of MaxPro while achieving statistical uniformity of the design-generation mechanism. Numerical experiments show unbiased Monte Carlo integration with reduced variance, excellent subspace projection performance, and competitive discrepancy properties. The methodology is further validated on benchmark engineering problems, including a meso-scale finite element model of concrete, demonstrating improved accuracy in surrogate modeling and probabilistic estimation.

The resulting criterion provides a simple and computationally efficient modification of MaxPro that enhances its robustness for nonadaptive computer experiments. The construction algorithm, open-source implementation, and reproducible optimized designs are provided to facilitate practical adoption of the method.

## 1. Introduction

Computer experiments are an indispensable tool in various contexts, including model screening, surrogate modeling, and numerical estimation of probabilistic integrals as part of uncertainty propagation analyses (encompassing statistical, sensitivity, and reliability assessments).

An important family of computer experiment designs is the space-filling designs, which aim to explore the entire design domain, including its boundaries, so that the simulation results adequately represent the space of all possible inputs. A seminal paper by Johnson et al.[13] introduced the Maximin and Minimax distance-based designs, which have since been generalized in Morris and Mitchell [25] and Pronzato [28] and elsewhere.

Another distinct class of space-filling designs is uniform designs, which strive to maintain a nearly uniform empirical distribution of design points over the domain. Their uniformity can be quantified by discrepancy measures; see Hickernell [11], which evaluate how evenly points are spread relative to an ideal uniform distribution.

* Corresponding author.
*Email address:* miroslav.vorechovsky@vut.cz (M. Vořechovský).

In both classes of designs, and in both model approximation and statistical integration settings, it can be crucial to have uniform and space-filling projections onto lower-dimensional subspaces. This is especially important in high-dimensional scenarios subject to the principle of sparsity, where only some subset of the factors may significantly affect the model output. When one knows (or strongly suspects) which factors are influential, it is therefore important to ensure that the corresponding lower-dimensional subspaces are well covered.

Good coverage in each one-dimensional projection can be achieved by restricting designs to Latin hypercube samples [2,21], and global space-filling properties can be attained by optimizing distance-based criteria (e.g., Maximin). However, ensuring good coverage simultaneously in all subspaces is more challenging; see Refs. [4,24,32] for related discussions.

To address this gap, Joseph et al. [14] proposed the Maximum Projection (MaxPro) criterion, which modifies the usual distance-based objective by emphasizing products of squared pairwise distances across coordinates. In effect, this encourages the design to be better spread not only in the full-dimensional space but also in its lower-dimensional projections. By treating the distance exponent or weights in a Bayesian manner, the mean (or expected) objective naturally features product terms in each subspace, thus favoring designs with good sub-dimensional coverage.

### 1.1. Research gap and contribution

Despite the extensive development of space-filling designs, two issues remain insufficiently addressed in the context of computer experiments. First, while the MaxPro criterion was specifically created to improve projection uniformity, its use of the standard Euclidean intersite distance leads to systematic density distortions near the boundaries of bounded domains. Second, no mechanism has been provided to guarantee statistical uniformity of the design-generation procedure, which is essential for unbiased integration and reliable performance in lower-dimensional subspaces. These observations identify a clear methodological gap: a projection-aware design that also achieves statistical uniformity.

The present work fills this gap by introducing a periodic variant of the MaxPro criterion (uMaxPro) that remedies boundary effects while preserving the projection structure of the original formulation. We build on the MaxPro criterion by introducing a simple enhancement that ensures *uniform* coverage of the design domain. We show that existing MaxPro designs systematically omit certain regions, reducing their effectiveness for some applications. By incorporating our enhancement, the resulting layouts not only remain space-filling in all subspaces, have an improved subspace behavior and competitive discrepancy properties across dimensions, but also attain a statistically uniform distribution of points. This dual property makes the enhanced designs suitable for both surrogate modeling (where a function approximation is needed) and *unbiased* Monte Carlo-type integration, thereby broadening the range of problems where MaxPro-based designs can be effectively employed.

### 1.2. The importance of sample uniformity

In this paper, two distinct notions of *uniformity* play an important role, and we state them explicitly to avoid confusion:

1. *Geometric (space–filling) uniformity of a deterministic design.* For a fixed set of points $D = \{x_s\}_{s=1}^{N_{\text{sim}}} \subset [0,1]^{N_{\text{var}}}$, one may evaluate how evenly these points fill the domain. This is a purely geometric property of a *single* design and can be quantified by discrepancy measures (e.g., the wrap–around discrepancy) or distance-based criteria such as Maximin [13]. No probability model is involved.
2. *Statistical (probabilistic) uniformity of a design–generating mechanism.* In practice, individual designs are often obtained by running a *stochastic* optimization procedure (random starts, random moves within simulated annealing, etc.). Thus the outcome is a *random design*. We call such a design–generating mechanism statistically uniform if, across many realizations of the design, each region of the domain is visited with probability proportional to its volume. Formally, if $X$ is a point drawn uniformly at random from a randomly generated design $D$, then statistical uniformity means $\mathbb{P}(X \in A) = |A|$ for any subregion $A \subseteq [0,1]^{N_{\text{var}}}$, where $|A|$ denotes its normalized Lebesgue measure.

Discrepancy therefore assesses geometric uniformity of a single deterministic design, whereas the histograms we construct later evaluate the statistical uniformity of the *design–generation procedure*. These two notions are complementary but not equivalent.

Monte Carlo simulation has long been the standard approach for numerical integration and uncertainty propagation in engineering and applied sciences [8,9,22,29]. In order to improve convergence properties and reduce estimator variance, quasi-Monte Carlo (QMC) methods based on low-discrepancy sequences were developed [1,17,26]. Randomized QMC techniques further combine variance reduction with unbiasedness of the estimator [27]. The present work is situated within this broader context, with a focus on the design of optimized deterministic point sets that preserve projection quality while ensuring statistical uniformity of the design-generation mechanism. Unlike classical QMC constructions, which are typically sequence-based, we consider optimization-based experimental designs tailored for computer experiments with fixed sample size.

In Monte Carlo numerical estimation, it is desirable to (a) minimize the number of integration points (thereby reducing computational cost), (b) reduce the variance of the estimated result, and (c) maintain an unbiased estimate that converges to the exact value. These conflicting demands are often addressed by optimizing the *sample uniformity* of the $N_{\text{sim}}$ integration points within the $N_{\text{var}}$-dimensional unit hypercube $[0,1]^{N_{\text{var}}}$. Furthermore, for an unbiased estimation, it is vital that the samples also exhibit *statistical uniformity*—that is, every region of the design domain is visited with the same probability.

Although one might intuitively suspect that sample uniformity influences the integration error, the relationship is indirect and somewhat vague. This connection is illustrated by the Koksma–Hlawka inequality [15], which bounds the integration error by the product of the function's variation (in the sense of Hardy and Krause) and the discrepancy of the sample points. In this sense, a lower discrepancy (indicating higher uniformity) reduces the upper bound on the error, though it does not directly quantify the actual error.

The impact of sample uniformity on the variance of the estimation becomes particularly pronounced when the scale of localized features of the integrated function (i.e., regions of high variation) is comparable to the minimum distance between points in a uniformly distributed sample; see, e.g., Section 5.1 of [36] and Section 6.5 of [20]. Such functions or metamodel surfaces do exist, however they do not represent the typical majority of engineering problems. For the bulk of the integrated or approximated functions, the scale of localized features dominates over the fluctuations in the uniformity of the point sample; see e.g. Liu and Wakeland [18]. In the majority of use cases, somewhat optimized point samples often work sufficiently as long as these are optimized enough to be free of point clusters that grossly violate uniformity.

Although reduced variance is desirable, it is essential that the estimator remains unbiased. Unbiasedness is a fundamental requirement for Monte Carlo integration: the sampling points should not statistically favor any particular region of the design domain, ensuring that every region is equally represented. This condition is particularly important when dealing with an unknown function or a *black-box* computer model. There are specific cases where it can be beneficial to oversample certain regions; for example, see Dette and Pepelyshev [3] concerning Gaussian process approximation. Similarly, importance sampling intentionally oversamples regions of higher contribution to the integral, but this approach requires a priori knowledge of the function's behavior. In the remainder of this work, we focus exclusively on the integration and approximation of an unknown function, without assuming any prior information about its properties.

We note that, in principle, a non-uniform sample can still yield an unbiased estimate if the sampling density is known and an appropriate importance-sampling correction is applied. However, in the context of optimized space-filling designs, the sampling distribution induced by the optimization procedure is not available in closed form and depends on the sample size, algorithmic details, and the specific local optimum reached. As a result, the effective density cannot be estimated from a single design, making importance-sampling corrections impractical. For this reason, statistical uniformity of the design-generation mechanism remains essential for obtaining unbiased Monte Carlo estimates in typical computer-experiment settings.

## 2. Distance-based space-filling designs

In 1990, Johnson et al. [13] published the possibly most widely used distance based space-filling criterion, the Maximin criterion. The Maximin criterion, $\phi_{\mathrm{Mm}}$, evaluates mutual distances between all pairs of points $i$ and $j$ within the sample and attempts to maximize the shortest distance among all pairs of points:

$$\phi_{\mathrm{Mm}} = \min_{\mathbf{x}_i, \mathbf{x}_j \in \mathcal{U}} d(\mathbf{x}_i, \mathbf{x}_j), \quad i \neq j. \tag{1}$$

The mutual distances are commonly measured by the Euclidean distance metric

$$d(\mathbf{x}_i, \mathbf{x}_j) = \sqrt{\sum_{v=1}^{N_{\mathrm{var}}} \left(\Delta_{ij,v}\right)^2}, \tag{2}$$

where the univariate projection distance of points $i, j$ onto dimension $v$ reads

$$\Delta_{ij,v} = |x_{i,v} - x_{j,v}|. \tag{3}$$

In 1995, Morris and Mitchell [25] proposed a modification of the Maximin criterion that incorporates the inverse distances between all pairs of points, rather than focusing solely on the pair with the smallest distance. For the Euclidean distance metric, their criterion is given by

$$\min_{D} \left\{ \frac{1}{\binom{N_{\mathrm{sim}}}{2}} \sum_{i=1}^{N_{\mathrm{sim}}-1} \sum_{j=i+1}^{N_{\mathrm{sim}}} \frac{1}{\left[\sum_{v=1}^{N_{\mathrm{var}}} \Delta_{ij,v}^2\right]^{k/2}} \right\}^{\frac{1}{k}}, \tag{4}$$

Moreover, in an unrestricted form, the Maximin criterion often favors point patterns with non-unique coordinate projections onto subspaces of lower dimensions (collapsible designs). If such a design is used for numerical integration of a function with weak or no interaction between the input variables, its performance is driven mainly by the 1D sampling uniformity, i.e., only by the number of unique coordinate projections. Morris and Mitchell [25] proposed a solution to this by restricting the possible solutions to the class of Latin Hypercube samples (LHS). The modified LH-Maximin criterion yielded designs with increased uniformity and performance in numerical integration.

Restricting poorly performing optimization criteria to LHS designs does improve the space-filling properties of the resulting point samples in two ways: (a) it restricts the ability of the optimization algorithm to form overly clustered points and (b) it automatically improves the projections onto 1D subspaces as these are firmly set as equidistant, i.e., $N_{\mathrm{sim}}$ coordinates have to be set into $N_{\mathrm{sim}}$ levels along each dimension. However, the restriction to LHS merely limits the possibilities of a faulty optimization criterion to fully manifest itself.

## 3. Maximum projection designs

### 3.1. The maximum projection criterion

When conducting a numerical experiment, it is known that commonly only a handful of the input variables (factors) affect the result significantly. This is called the *sparsity effect principle*. If these governing input variables are identified, the dimensionality (complexity) of the problem at hand can be reduced drastically to a fraction of the original $N_{\mathrm{var}}$. The subsequent computational effort is therefore incomparably more effective. However, there remains a risk of missing important interactions among factors or not incorporating potentially critical input variables at all. Therefore the design domain of input variables may contain a number of redundant variables and the sparsity effect can be used only after recognizing the key influencing factors e.g., by a sensitivity analysis done upon the results of initial screening. Until then, the point samples should treat all input variables with an equal importance and attempt to sample all contained subspaces (combinations of factors) as uniformly as possible.

It is true that an LH-Maximin criterion improves space-filling in the full $N_{\mathrm{var}}$-dimensional space and ensures uniformity in each one-dimensional coordinate projection. However, it does not inherently guarantee uniformity in the remaining $(N_{\mathrm{var}} - 1)$- to 2-dimensional subspaces. In order to improve the uniformity in the subspaces, in 2015, Joseph et al. [14] defined in their Eq. (5) the Maximum Projection design criterion (further referred to as MaxPro) as

$$\min_{D} \left\{ \frac{1}{\binom{N_{\mathrm{sim}}}{2}} \sum_{i=1}^{N_{\mathrm{sim}}-1} \sum_{j=i+1}^{N_{\mathrm{sim}}} \frac{1}{\prod_{v=1}^{N_{\mathrm{var}}} (x_{i,v} - x_{j,v})^2} \right\}^{\frac{1}{N_{\mathrm{var}}}}. \tag{5}$$

The denominator features a *product* of all $N_{\mathrm{var}}$ squared 1D distance projections for each pair of points, while the relaxed Maximin criterion with Euclidean metric has their *sum*; recall Eq. (4). Therefore, the MaxPro criterion prefers designs that maximize the point coordinate projections onto all subspaces. Here, no pair of points can get too close in any of the projections. Therefore, unlike Maximin, the MaxPro criterion penalizes point clustering in all available subspaces. Thus, one may anticipate that a MaxPro-optimal design would yield a uniform distribution of points within any subspace. However, this is not the case.

### 3.2. Statistical (non)uniformity of maximum projection designs

To study the *statistical* behavior of the designs, we repeatedly optimize the MaxPro criterion using random initializations and collect the resulting designs to construct histograms in the design domain. All designs used in the histogram analyses are generated as follows. For each realization, we draw an independent random Latin hypercube design and apply a simulated annealing (SA) optimization [35] (see below) based on coordinate–swap moves until the SA termination criterion is reached. This produces one locally optimal MaxPro design. The procedure is then repeated independently $N_{\mathrm{run}} = 10^4$ times, yielding $N_{\mathrm{run}}$ separate optimized LHS designs.

We inspect the statistical uniformity of the point distribution across the design domain. The design domain is divided into $n_b = {N_{\mathrm{sim}}}^{N_{\mathrm{var}}}$ bins (much like the LH bins). For each bin, we count the number of point occurrences across the $N_{\mathrm{run}}$ MaxPro sample realizations. For statistically uniform point samples, the average frequency of each bin being visited by a point is $f_{\mathrm{u}} = N_{\mathrm{sim}} N_{\mathrm{run}} / n_b$. In the analysis of MaxPro samples, we divide the respective bin frequencies by this "uniform" frequency to obtain the relative frequency $f$ providing the sense of over- or under-sampling of individual bins. For truly uniform samples, all bins must have an equal probability of being visited by a point, i.e., an equal, unit frequency of point occurrence. In other words, if the design-generating mechanism were statistically uniform, then the expected bin frequencies would be proportional to the bin volumes, and the histogram would be approximately flat.

Fig. 1 shows that this is not the case for MaxPro: certain regions, especially near domain corners, are visited with systematically lower frequency, while regions away from boundaries are oversampled. In this probabilistic sense, the MaxPro mechanism fails to satisfy statistical uniformity. The top row of Fig. 1 shows histograms of all 2D subspaces of design domains of dimensions from $N_{\text{var}} = 2$ to $N_{\text{var}} = 5$ containing MaxPro samples of 32 points ($N_{\text{sim}}$). The bottom row of Fig. 1 attempts to display the full spatial distribution of points within $N_{\text{var}} = 3$ and $N_{\text{var}} = 4$ dimensional domains. In the case of a $N_{\text{var}} = 3$ domain, 2D histograms are constructed for $N_{\text{sim}}$ layers (LHS levels – increments) along the third dimension in a consecutive order. For the $N_{\text{var}} = 4$ domain, such an approach is looped also over the fourth dimension.

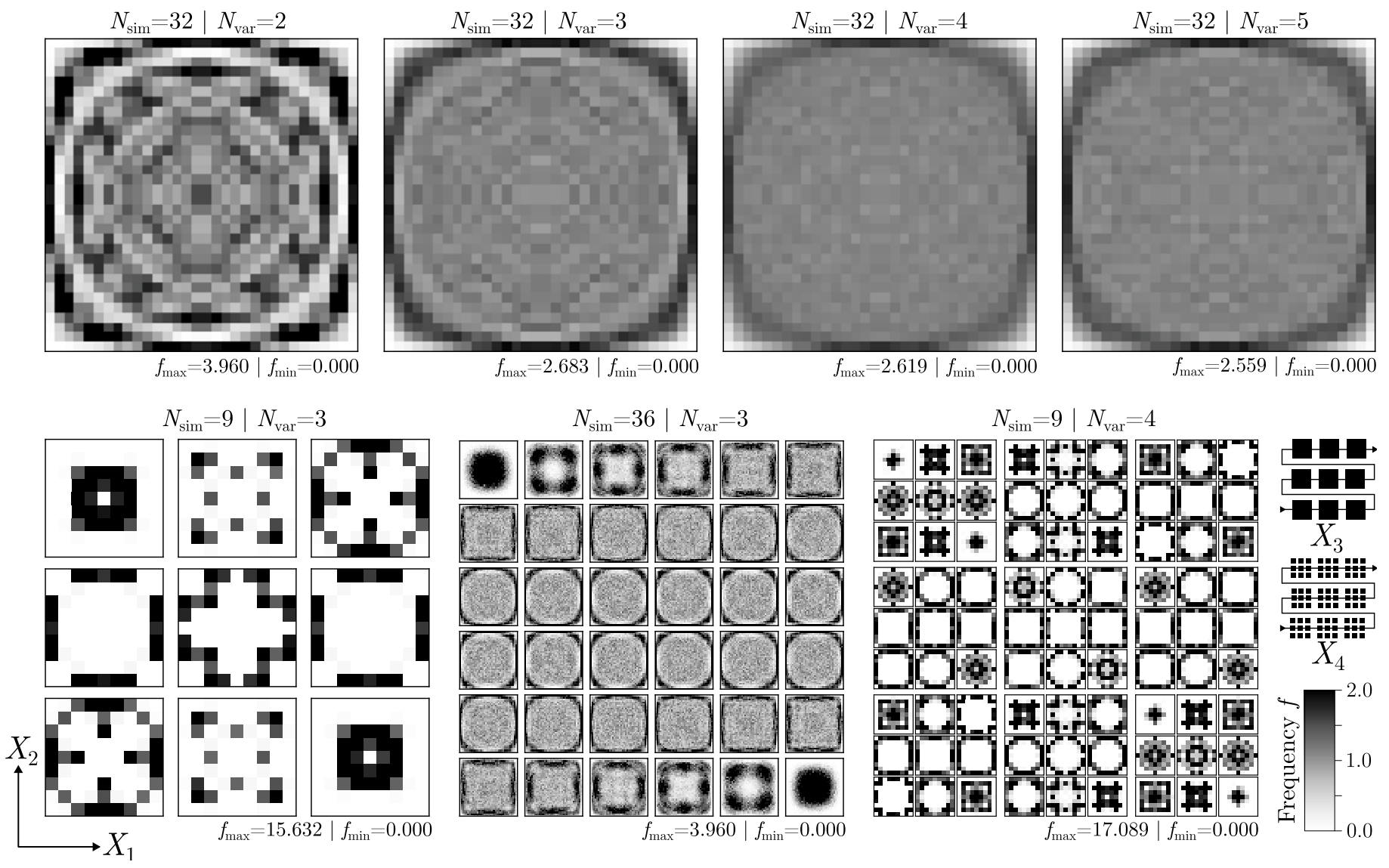

**Fig. 1.** Histograms of relative frequencies in which the MaxPro criterion visits given LH bins for various sample sizes, $N_{\text{sim}}$, and domain dimensions, $N_{\text{var}}$. The frequencies are normalized so that a value of 1 corresponds to perfect statistical uniformity (each bin is visited with probability proportional to its volume). Values below 1 indicate systematically undersampled regions, and values above 1 indicate oversampled regions. These histograms summarize the empirical distribution obtained from $N_{\text{run}}$ independent optimized designs.

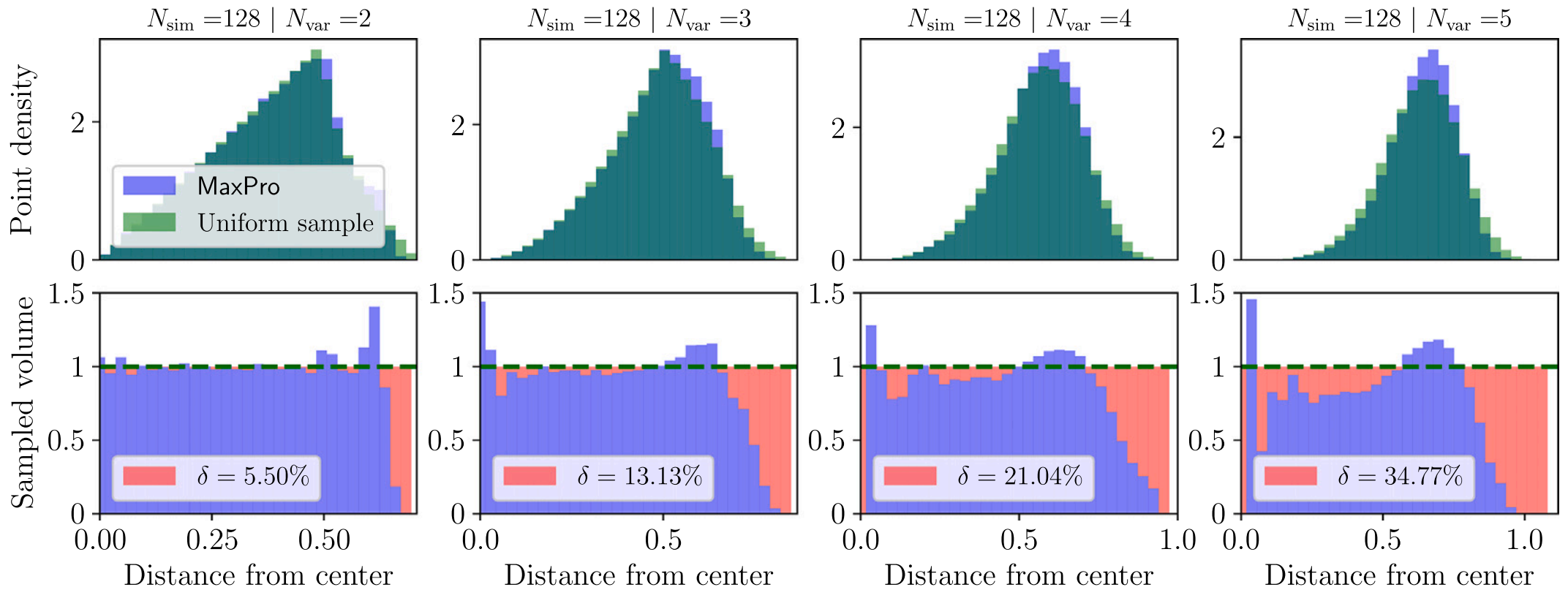

**Fig. 2.** Radial histograms of MaxPro 128-point samples within 2D, 3D, 4D and 5D design domains. Deviations from the reference uniform profile highlight the systematic undersampling of corner regions by MaxPro, which becomes more pronounced in higher dimensions.

It can be seen that the original MaxPro criterion produces point samples that *never* visit a major portion of the domain. This behavior inevitably leads to biased estimates when using these point samples for Monte Carlo integration or approximation. The MaxPro criterion avoids sampling of domain corner regions while creating an oversampled layer farther from boundaries. With an increasing domain dimension, $N_{\text{var}}$, it appears that the problem is vanishing. However, this conclusion is substantially misleading. With increasing dimension, the corners account for nearly all of the hypercube volume; see Section 6 in [34].

We now examine the point distribution in greater detail. To this end, we construct histograms of the point locations with respect to their distance from the hypercube center; see the first row of Fig. 2. We plot the histograms for the LH–MaxPro samples and, for comparison, for an ideally uniform point sample. Certain deviations between the two can be seen, although these still do not seem critical. However, in the second row of Fig. 2, the histograms are weighted by the portion of domain volume contained in the respective regions (radial layers/bins). This volume-corrected view reveals the undersampling of corner regions caused by the MaxPro metric. We also estimate the portion of the normalized undersampled volume, $\delta$: the deficit of the sampling density relative to the uniform distribution. The total red-shaded area represents the measure of volume in each radial layer in which the optimized design is less dense than a uniform distribution. Thus, $\delta$ measures the fraction of the domain volume in which the optimized design is less dense than a uniform reference. As seen in Fig. 2, this undersampling becomes more pronounced as the dimension of the design domain increases.

The MaxPro criterion fails to satisfy the fundamental requirement of statistical uniformity—that is, it does not select all regions of the design domain with equal probability. This systematic bias in the point distribution inevitably leads to estimation bias in both numerical integration and approximation.

Joseph et al. [14] were aware of the nonuniformity issue and observed that "both maximum projection and generalized maximin Latin hypercube designs favor more points towards the boundaries than at the center, which affects the overall uniformity of the points in the design space." They further noted that, surprisingly, although Latin hypercube designs impose restrictions on the equal spacing of levels, their combination with the MaxPro criterion improves the uniformity in all subspaces. Nonetheless, they dismissed this nonuniformity by arguing that the primary objective of a computer experiment is function approximation rather than integration, and that "the poor performance of maximum projection designs under the uniformity measure is not of great concern."

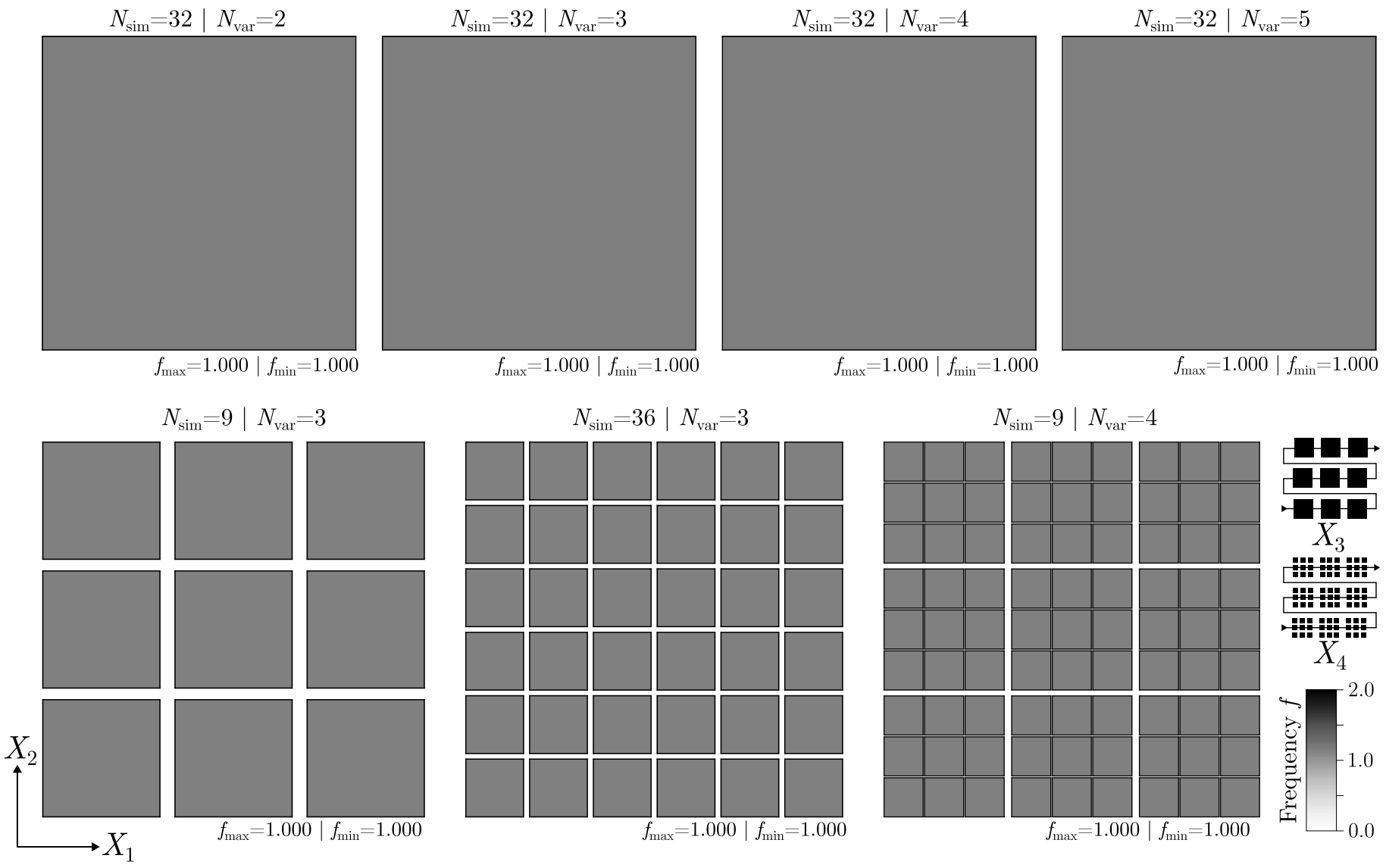

**Fig. 3.** Histograms of relative frequencies in which the proposed uMaxPro criterion visits given LH bins for various sample sizes, $N_{\mathrm{sim}}$, and domain dimensions, $N_{\mathrm{var}}$. The frequencies are normalized so that a value of 1 corresponds to perfect statistical uniformity (each bin visited with probability proportional to its volume). Values below 1 indicate systematically undersampled regions, and values above 1 indicate oversampled regions. These histograms summarize the empirical distribution obtained from $N_{\mathrm{run}}$ independent optimized designs.

In what follows, we present a critical improvement of the original MaxPro criterion that achieves its intended goal: to combine excellent projection properties with complete statistical uniformity.

## 4. Uniform maximum projection designs

The problem of the MaxPro criterion unintentionally oversampling the boundary regions is connected to the selection of the distance metric used. The original work proposes using the *intersite* Euclidean distance metric, see Eq. (2), much like the Maximin criterion. Inevitably however, this leads to an uneven distribution of directions of pairwise distance vectors, especially towards the boundary. Therefore, in line with our previous improvement of the class of distance-based space-filling designs [33], we propose to enhance the projection-based MaxPro designs by redefining the metric. In particular, we propose measuring the mutual pairwise distance not simply as the direct distance between points $i$ and $j$, but by employing the *minimum image convention* to compute the distance between their closest periodic images within a borderless, periodically repeated design domain. In this approach, the distance $\Delta_{ij,v}$ from Eq. (3) is replaced by its periodic variant $\overline{\Delta}_{ij,v}$

$$\overline{\Delta}_{ij,v} = \min\left(\Delta_{ij,v},\ 1 - \Delta_{ij,v}\right). \tag{6}$$

As a result, we introduce a periodic variant of the Maximum Projection criterion—referred to as the *uniform* Maximum Projection criterion (hereafter, uMaxPro)—which is to be minimized:

$$\overline{\psi}(D) = \left\{ \frac{1}{\binom{N_{\mathrm{sim}}}{2}} \sum_{i=1}^{N_{\mathrm{sim}}-1} \sum_{j=i+1}^{N_{\mathrm{sim}}} \frac{1}{\prod_{v=1}^{N_{\mathrm{var}}} \overline{\Delta}_{ij,v}^{\,2}} \right\}^{\frac{1}{N_{\mathrm{var}}}} \tag{7}$$

Unlike the original MaxPro design–generation mechanism, the periodic variant uMaxPro does not favor any particular region of the domain. Because the periodic distance $\overline{\Delta}_{ij,v}$ is invariant under translations of the coordinate system, the induced sampling distribution becomes essentially translation–invariant as well. In repeated realizations, each bin (in the full space and in all coordinate projections) is visited with frequency proportional to its volume, see the histograms in Fig. 3. Thus the periodic MaxPro mechanism achieves statistical uniformity while maintaining the strong geometric uniformity and projection structure of MaxPro in each deterministic realization. In this way, the periodicity remedies the issues associated with domain boundaries.

This translation invariance implies that no region of the design domain is preferred by the optimization criterion. Therefore, even if individual realizations are not perfectly (geometrically) uniform, repeated independent realizations combined with arbitrary global translations induce a statistically uniform sampling mechanism in the sense that each subregion is visited with probability proportional to its volume. This statistical uniformity is not a property of a single deterministic realization, but rather a property of the distribution induced by the design-generation mechanism. It follows from the group symmetry of the criterion, specifically its translation invariance modulo one.

A possible concern is that applying the minimum operation in Eq. (6) along individual dimensions is not rotationally invariant. However, although the Euclidean distance metric is invariant under rotations, the original MaxPro design criterion is not. Thus, introducing the periodic metric does not compromise the overall design, and any loss of rotational invariance is inconsequential.

The resulting point samples can be compared by evaluating the weighted histograms presented in Figs. 2 and 4. The uMaxPro criterion does not tend to over- or under-sample any domain regions. This property leads to an unbiased estimation with a reduced variance while the projection properties are already optimized as an inherent feature of the MaxPro criterion. From an implementation standpoint, the periodic modification requires only a minor adjustment to existing code, and the computational overhead of evaluating Eq. (6) is negligible.

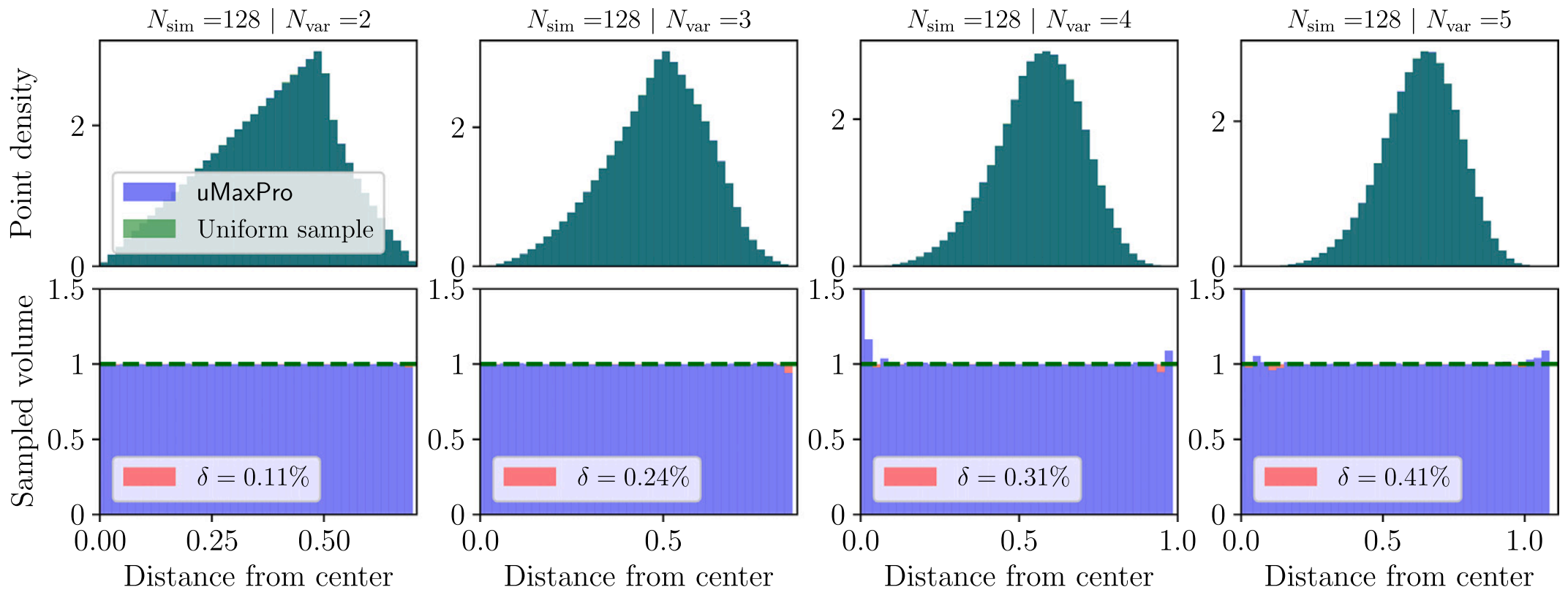

**Fig. 4.** Histograms of radial distribution of uMaxPro 128-point samples within 2D, 3D, 4D and 5D design domains. The uMaxPro histograms match the uniform reference, confirming the elimination of boundary bias.

## 5. Optimized sample construction algorithm

To obtain point samples optimized according to the introduced uMaxPro criterion, we employ a simulated annealing (SA) algorithm, following the framework previously used for correlation-optimal Latin hypercubes [35]. All optimized designs considered in this paper are restricted to the class of Latin hypercube samples. The simulated annealing algorithm itself is not limited to LHS structures; however, in the present study it operates within the LHS framework to preserve marginal stratification. SA is particularly suitable for design problems with discrete marginals, such as LHS designs, because it naturally operates on coordinate-swap moves within a fixed grid of levels.

The optimization problem is combinatorial and highly multimodal, and therefore a deterministic algorithm would typically become trapped in poor local minima. Simulated annealing offers a practical compromise: while it cannot guarantee a global optimum, its probabilistic acceptance rule enables occasional uphill moves that help escape local minima and explore promising regions of the design space. Throughout the annealing schedule we record the best design encountered, and after the temperature reaches its minimum we apply a short greedy (local-improvement) phase to ensure that the stored design is at least locally optimal with respect to all single-coordinate swaps.

This procedure has proven effective in practice, consistently producing designs that are near-optimal under both the MaxPro and uMaxPro criteria. The implementation is provided in the open-source Python package `uMaxPro` [38], which was used to generate all designs and histograms reported in this paper. Reproducible optimized designs for both the original MaxPro (using the intersite metric) and the proposed uMaxPro (periodic metric) are available in the accompanying `zenodo` repository [37].

## 6. Numerical studies

In what follows, we present numerical studies comparing properties of point samples optimized according to the original MaxPro criterion and the proposed uMaxPro criterion.

### 6.1. Discrepancy

As a standardized measure of the experimental design uniformity can be considered the discrepancy. To avoid problems regarding domain boundaries, we select the Wrap-around discrepancy ($\mathrm{WD}_2$) proposed by Hickernell [10]; see also Fang and Ma [7]. The $\mathrm{WD}_2$ discrepancy is insensitive to the order of sampling points, swapping of point coordinates between dimensions (rotation-invariant), and mirroring of point coordinates (reflection-invariant). Another truly unique property of the $\mathrm{WD}_2$ discrepancy is its invariance to coordinate shifts (unanchored discrepancy), i.e., it does not anchor discrepancy values with respect to e.g., the domain origin or domain center. Therefore, the $\mathrm{WD}_2$ discrepancy is not affected by the domain boundaries whatsoever. The expression for the $\mathrm{WD}_2$ discrepancy reads

$$[\mathrm{WD}_2(\mathscr{D})]^2 = -\left(\frac{4}{3}\right)^{N_{var}} + \frac{1}{N_{sim}}\left(\frac{3}{2}\right)^{N_{var}} + \frac{2}{N_{sim}^2}\sum_{i=1}^{N_{sim}-1}\sum_{j=i+1}^{N_{sim}}\prod_{v=1}^{N_{var}}\left[\frac{3}{2} - \Delta_{ij,v}\left(1-\Delta_{ij,v}\right)\right], \tag{8}$$

where $\Delta_{ij,v}$ is the projection distance from Eq. (3). The product $\Delta_{ij,v}\left(1-\Delta_{ij,v}\right)$ removes the effect of domain boundaries; recall the minimum image convention metric in Eq. (6).

Due to evaluating all mutual 1D distance projections, the $\mathrm{WD}_2$ discrepancy exhibits a high sensitivity to the uniformity of coordinates along each marginal. Any clustered pair of coordinates is detrimental to the $\mathrm{WD}_2$ value.

Although LH sampling enforces marginal stratification, the wrap-around discrepancy remains sensitive to the relative 1D distances between projected coordinates. Therefore, even within the class of LHS designs, different within-stratum placements may lead to different discrepancy values.

In the present study, all LH designs are generated using the LHS-median (midpoint) placement within each stratum, following the terminology introduced in Vořechovský and Novák [35]. In that work, three variants of Latin hypercube construction were distinguished: LHS-random (uniform sampling within each stratum), LHS-median (sampling at the midpoint of each stratum), and LHS-mean (sampling at the mean of the conditional distribution restricted to the stratum). For the uniform unit hypercube considered here, the LHS-median and LHS-mean schemes coincide. As a consequence, no additional within-stratum randomness is present in the one-dimensional projections. The discrepancy values therefore reflect purely the structural properties of the optimized designs rather than additional perturbations induced by LHS-random sampling. This explains why no within-stratum dispersion effects are visible in Fig. 5.

The Halton sequence is specifically designed to minimize discrepancy, achieving one of the steepest rates of decrease and thus can be regarded as nearly optimal in this respect. Fig. 5 shows the wrap-around discrepancy as a function of the sample size for dimensions $N_{var}$ = 2–5. The numerical results demonstrate that both the original MaxPro and the uniform Maximum Projection (uMaxPro) designs exhibit a decrease rate comparable to the Halton sequence, attaining a rate of $\mathcal{O}(1/N_{sim}^2)$ for $[\mathrm{WD}_2]^2$ (i.e., $\mathcal{O}(1/N_{sim})$ for $\mathrm{WD}_2$) in two dimensions. Although the absolute rate becomes slower in higher dimensions, the uMaxPro designs maintain discrepancy values close to those of the Halton sequence across all tested dimensions.

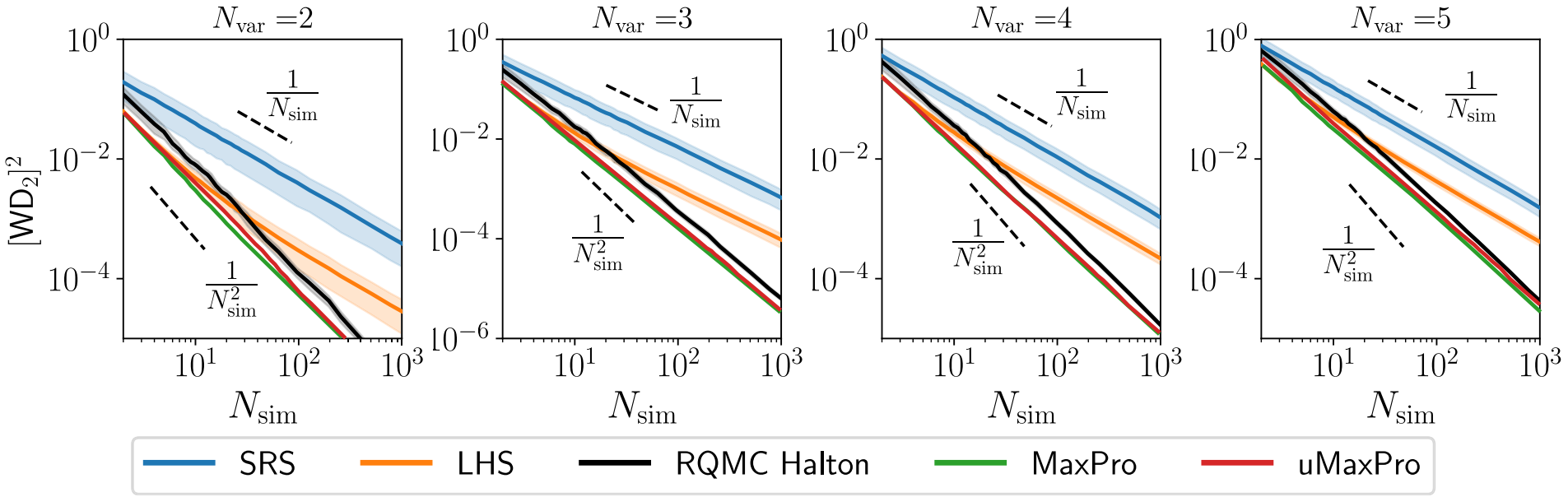

**Fig. 5.** Discrepancy comparison of 2D, 3D, 4D and 5D design domains.

It is noteworthy that the discrepancy measure is only indirectly connected to the expected performance in numerical integration or approximation. Therefore, the sample performance in discrepancy should not be understood as a direct measure of "uniformity" as such. The connection to the intuitive notion of uniformity is not straightforward. Although both the MaxPro and uMaxPro criteria yield excellent discrepancy values, their actual performance in Monte Carlo integration is to be revealed later.

### 6.2. Periodic maximin criterion

The Periodic Maximin criterion [33], $\phi_{PMm}$, is a modification of the original Maximin criterion that utilizes the minimum image convention. The criterion evaluates mutual periodic distances between all pairs of points $i$ and $j$ within the sample and attempts to maximize the shortest distance among all pairs of points:

$$\phi_{PMm} = \min_{\mathbf{x}_i,\mathbf{x}_j \in \mathcal{U}} \overline{d}(\mathbf{x}_i, \mathbf{x}_j), \quad i \neq j. \tag{9}$$

The mutual distances are here measured by the periodic Euclidean distance metric

$$\overline{d}(\mathbf{x}_i, \mathbf{x}_j) = \sqrt{\sum_{v=1}^{N_{var}} \left(\overline{\Delta}_{ij,v}\right)^2} \tag{10}$$

with the periodic projection distance $\overline{\Delta}_{ij,v}$ defined in Eq. (6).

The results displayed in Fig. 6 show that the proposed uMaxPro yields the best (higher is better) values of the shortest pairwise distance found in the sample out of all compared sampling methods. Both LH-MaxPro and LH-uMaxPro designs outperform even the Halton sequence, let alone the simple random samples and LH designs with random ordering. The $\phi_{PMm}$ values of uMaxPro designs show a systematic improvement over the original MaxPro designs that already performed well in the Maximin benchmark in Joseph et al. [14].

### 6.3. Integrating functions with strong variable interactions

The overall sample uniformity and projection properties can be conveniently tested using a function that exhibits strong interactions among input random variables. Let us consider the multivariate function of a vector of independent standard normal variables, $X_v$, defined as

$$g_{\Pi exp}(\boldsymbol{X}) = \prod_{v=1}^{N_{var}} \exp(-X_v^2). \tag{11}$$

The exact mean value of the function reads; see Appendix A in Mašek and Vořechovský [20]

$$\mathsf{E}[g_{\Pi exp}] \equiv I = \frac{1}{\sqrt{3^{N_{var}}}}. \tag{12}$$

Monte Carlo integration estimates the solution by averaging, assigning an equal weight of $1/N_{sim}$ to all $N_{sim}$ integration points

$$I_N = \frac{1}{N_{sim}} \sum_{s=1}^{N_{sim}} g_{\Pi exp}(\boldsymbol{X}_s). \tag{13}$$

Given $N_{run}$ realizations of the sample, each comprising $N_{sim}$ points and providing an estimate $I_N$, we obtain $N_{run}$ integral estimates. The RMSE is then approximated by calculating the average of these estimates

$$\mathsf{RMSE}[I_N] \approx \sqrt{\frac{1}{N_{run}} \sum_{r=1}^{N_{run}} \left(I_{N,r} - I\right)^2}. \tag{14}$$

Now, let us estimate the mean value, $\mathsf{E}[g_{\Pi exp}]$, and report the root mean square error $\mathsf{RMSE}[g_{\Pi exp}]$, as obtained by $N_{run} = 500$ independent runs. Samples obtained by various sampling methods in dimensions $N_{var} = 2, 3, 4, 5$ are used. Sampling points are transformed from a unit hypercube into standard Gaussian variables by the univariate inverse cumulative distribution functions.

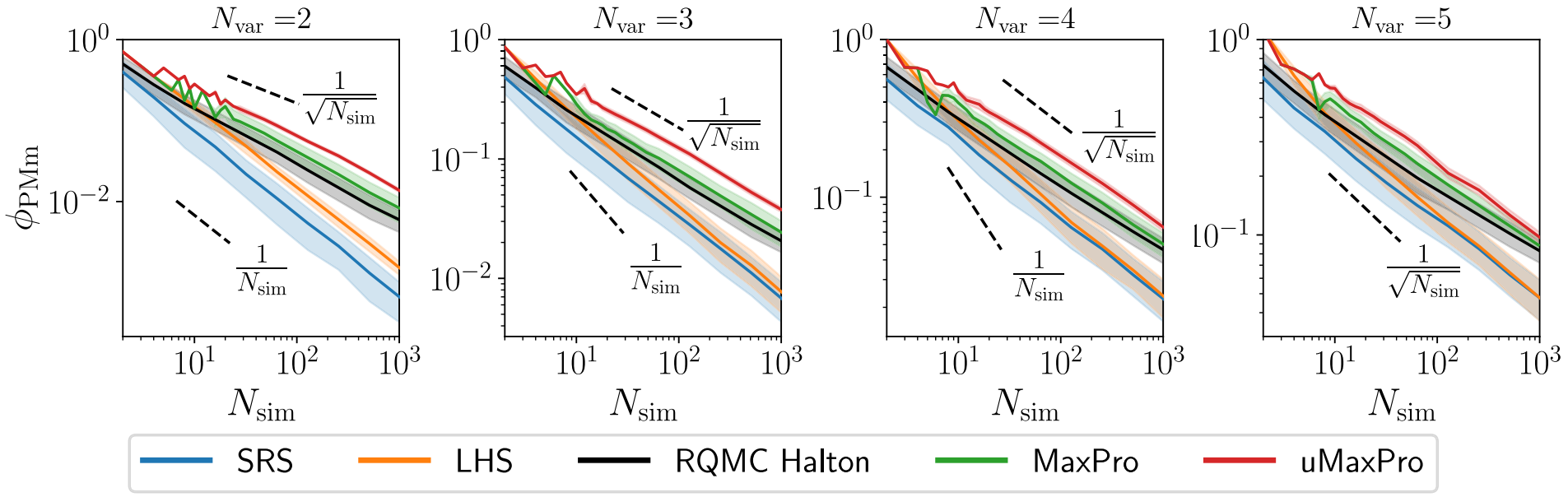

**Fig. 6.** Periodic Maximin values comparison in 2D, 3D, 4D and 5D design domains (higher is better).

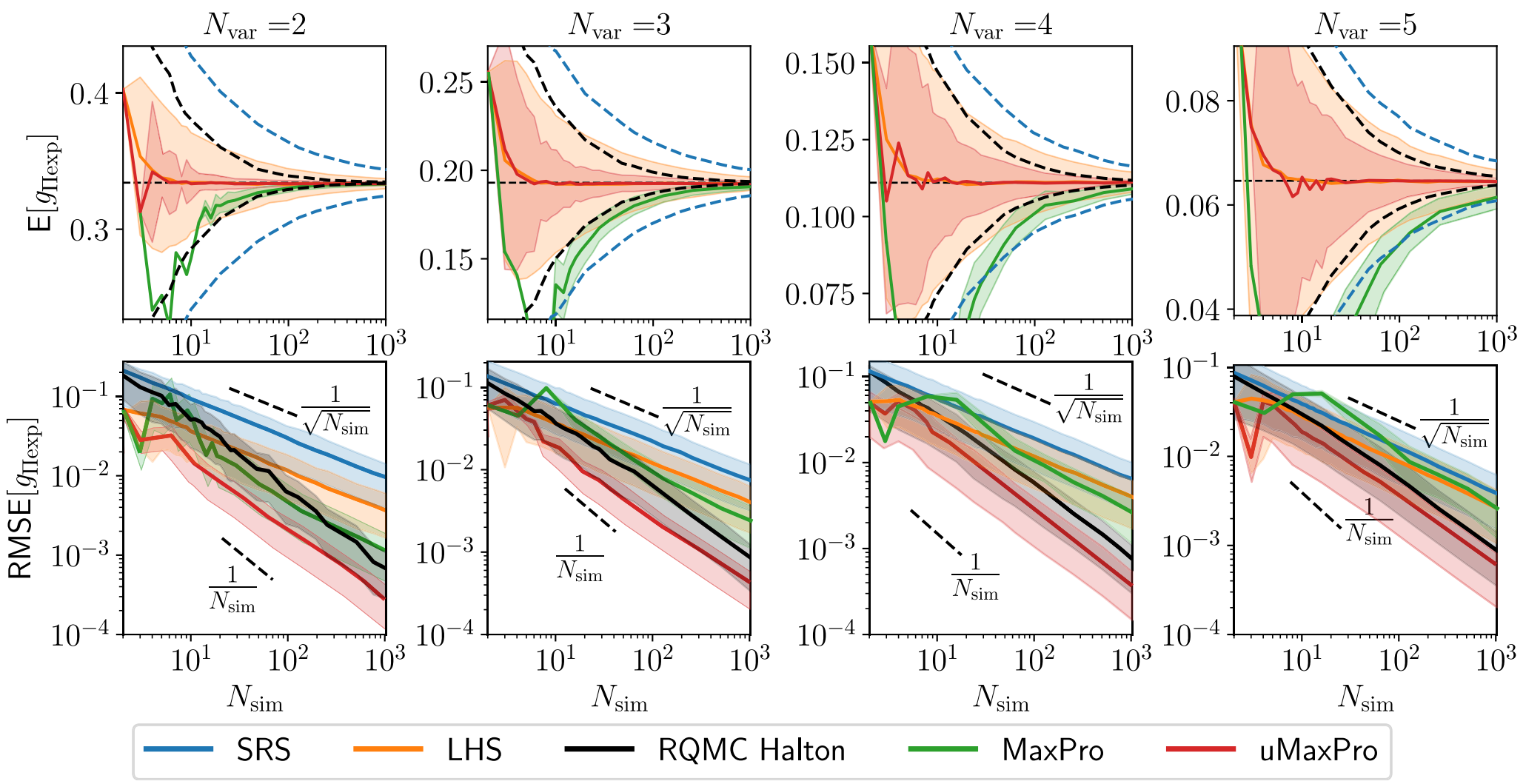

**Fig. 7.** Convergence comparison of the estimated mean value and the variance of the estimator for a function of $N_{var} = 2, 3, 4, 5$ strongly interacting variables. Methods yielding unbiased estimates have their $\pm\sigma$ bands plotted with dashed edges; the biased method (MaxPro) shows a clear separation between its estimated mean curve and the true value (horizontal lines in the top row). The bottom row displays the RMSE, where the steeply descending curves for uMaxPro consistently indicate the lowest errors across all dimensions.

The top subplot of Fig. 7 compares the convergence plots for the mean value estimations. The estimated mean value is shown along with a band representing $\pm$ one standard deviation ($\sigma$). To enhance clarity, methods providing unbiased estimations are represented using dashed edges for the $\pm\,\sigma$ band. In the bottom subplot of Fig. 7, the root mean square error RMSE$[g_{\Pi exp}]$ is depicted. The results indicate that the proposed uMaxPro criterion provides substantially improved estimation accuracy compared with the original MaxPro criterion along with an unbiased estimation due to statistical sample uniformity.

The uMaxPro rate of decrease of the root mean square error is close to $1/N_{sim}$, comparable to the Halton sequence. However, the uMaxPro yields an error systematically smaller than the Halton sequence.

### 6.4. Subspace projection benchmark

The MaxPro criterion was originally proposed as a sample optimization criterion that maximizes space-filling properties not only within the full domain dimension but also in all subsets of factors. This is important if only some of the factors are active.

We now investigate the projection properties of the selected sampling methods within a 5D design domain ($N_{var} = 5$) across all possible subspaces of dimensions $N_{var,sub} = 2, 3, 4 < N_{var}$. We estimate $N_{var,sub}$-dimensional variants of the discrepancy (Section 6.1) and the product function (Section 6.3).

In Fig. 8, displayed are the results of the discrepancy analysis in all possible subspace combinations. It can be concluded that the original MaxPro criterion retains an excellent discrepancy even in subspace projections, as was its original goal. The proposed uMaxPro criterion exhibits nearly identical discrepancy, which demonstrates that sample discrepancy is not affected by the periodic modification.

Fig. 9 displays the estimated mean value of the $g_{\Pi exp}$ function, recall (Section 6.3), and RMSE of the estimation in all possible subspace combinations. Similarly to the analysis in full dimension, recall Section 6.3, LH-MaxPro designs provide a biased estimation which makes them unusable in Monte Carlo integration. However, in the subspace analysis, MaxPro criterion performs even worse than randomly ordered LH samples. This is due to the statistical non-uniformity of MaxPro samples, i.e. undersampled domain corners and oversampled boundaries. When projecting such a nonuniform sample onto subspaces, the under- or oversampled regions tend to further accumulate and the sample nonuniformity increases as the sample is projected onto increasingly lower subspaces.

The periodic modification, uMaxPro, is a crucial improvement over the original MaxPro criterion that allows one to reach the original goal, i.e., to perform strongly when integrating a function, especially within subspaces of the original point sample. The uMaxPro criterion, on the other hand, retains an unbiased estimation along with strong variance reduction. In the subspaces, the uMaxPro samples continue to exhibit strong performance, clearly surpassing even the Halton sequence.

## 7. Application to engineering problems

This section illustrates how the proposed uMaxPro criterion, the associated software implementation, and the precomputed designs can be employed in practical engineering settings. We consider two benchmark problems from structural mechanics: a short column in ultimate limit state and a cantilever beam in serviceability limit state. For each problem, the designs are used to generate input samples, the corresponding responses are evaluated by the underlying computer model, and a Gaussian Process (Kriging) surrogate is constructed. The quality of the surrogate is then quantified in terms of the prediction error, allowing a direct comparison of the competing design strategies.

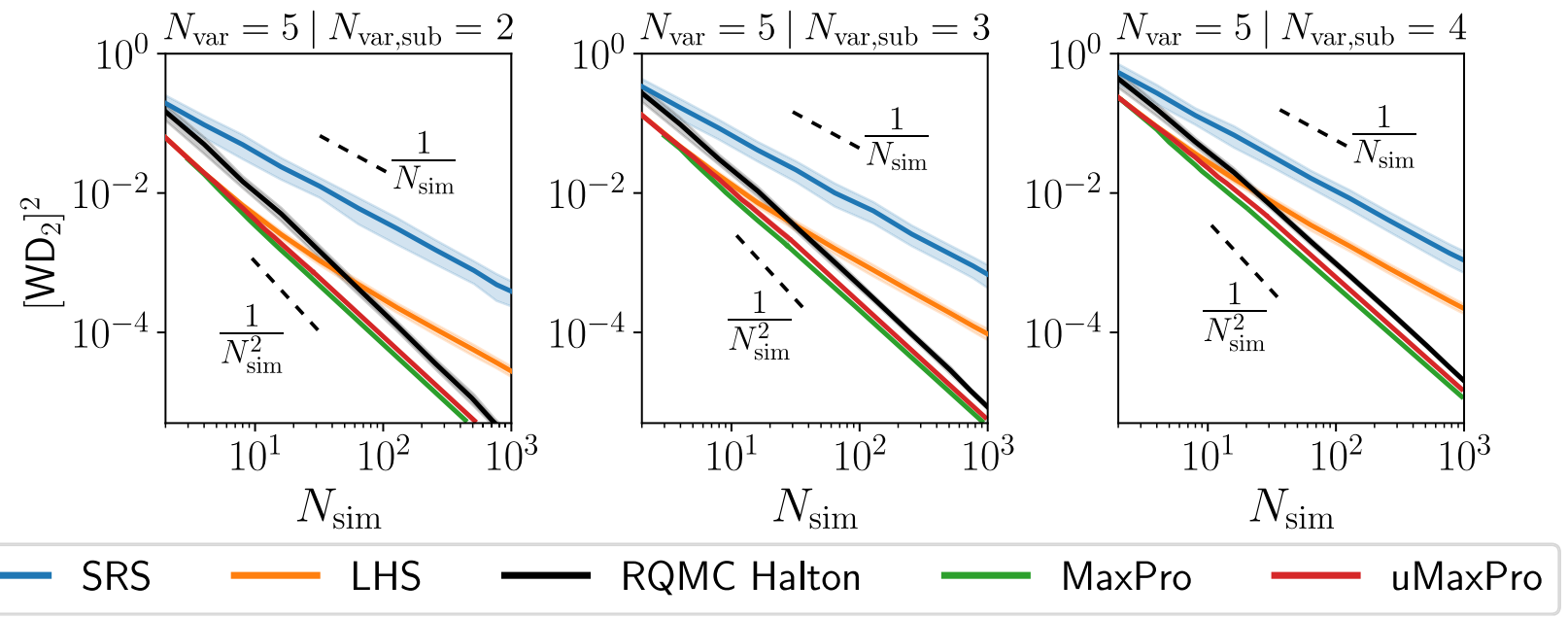

**Fig. 8.** Discrepancy of 5D designs computed for all 2D, 3D and 4D subspaces.

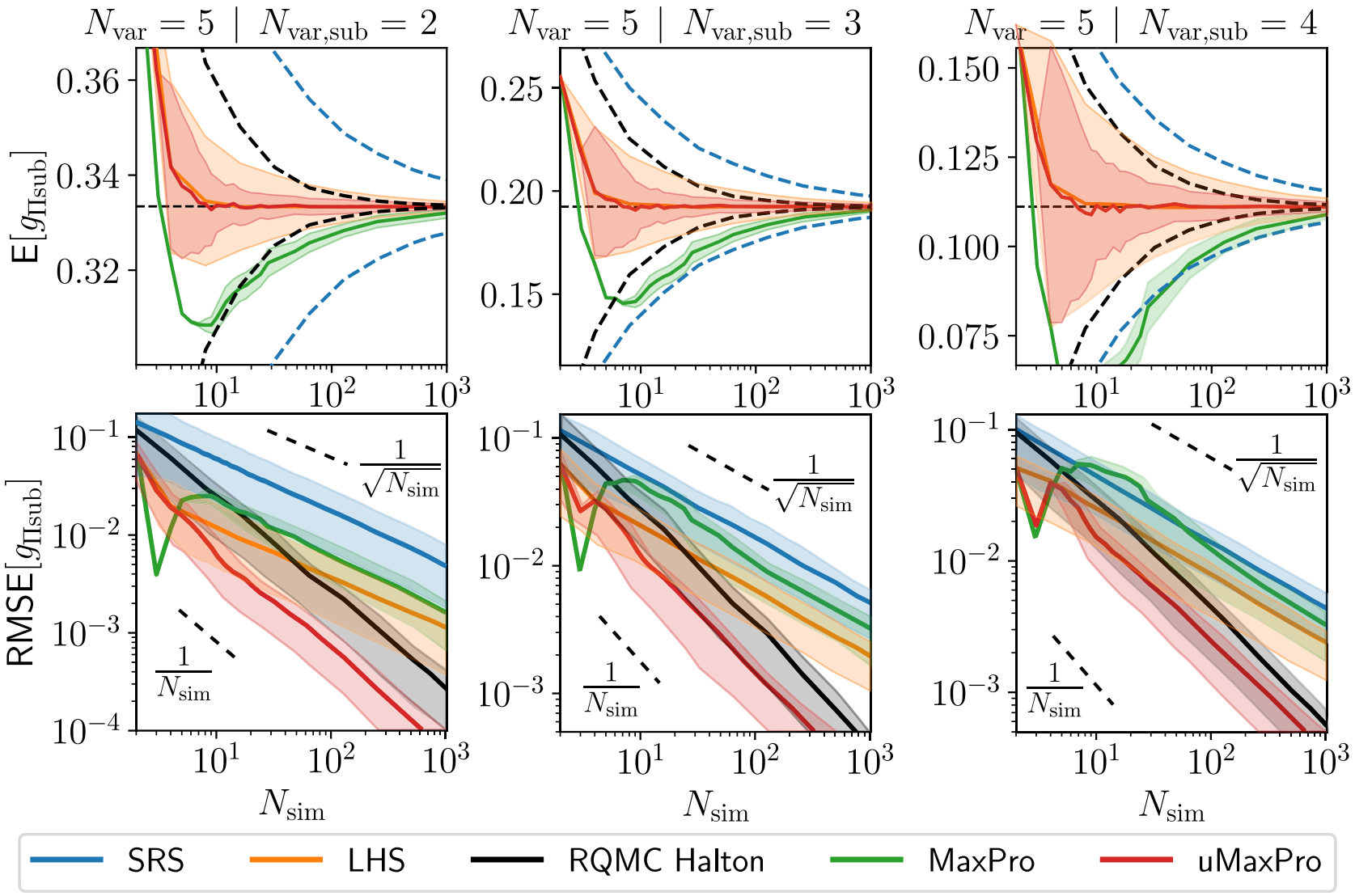

**Fig. 9.** Product function comparison in 2D, 3D and 4D subspaces of a 5D design domain. The shaded bands represent ± one standard deviation.

### 7.1. Benchmark problems and projection settings

The two examples are taken from the *Virtual Library of Simulation Experiments: Test Functions and Datasets* [31], which is a widely used repository for benchmarking methods in the design and analysis of computer experiments.

In the context of projection properties, we focus on a challenging setting in which the response depends on one dimension fewer than the optimized design. More precisely, we argue that the $N_{\text{var}} = D - 1$ dimensional subspace represents a demanding scenario when analyzing an $N_{\text{var}} = D$ dimensional point sample.

In order to deliberately test projection robustness, the response functions are evaluated in a lower-dimensional subspace than the dimension in which the design is optimized. More precisely, we construct the sampling design in $D$ dimensions, while the underlying engineering model depends only on $D - 1$ variables.

The input samples used in the following examples are therefore taken from $N_{\text{var}} = 3$ dimensional subspaces of $N_{\text{var}} = 4$ dimensional designs for the short column problem (see Section 7.2), and from $N_{\text{var}} = 4$ dimensional subspaces of $N_{\text{var}} = 5$ dimensional designs for the cantilever beam problem (see Section 7.3). In each case, one additional design dimension is included as a redundant variable in the original design and then omitted at the stage of model evaluation.

This setup emulates the common practical situation in which the active input variables are not known a priori. In real computer experiments, some variables may turn out to be inactive or weakly influential only after screening or sensitivity analysis. A projection-robust design should therefore retain good space-filling and uniformity properties even after projection onto arbitrary lower-dimensional subspaces.

The redundant dimensions in the examples below are intentionally retained during design construction and removed only at the model-evaluation stage. This provides a controlled and stringent test of the projection behavior of the competing sampling criteria.

**Table 1**
Input random variables and their distributions.

| Variable | Distribution | Mean ($\mu$) | Std. Dev. ($\sigma$) |
|---|---|---|---|
| $Y$ | Lognormal | 5 | 0.5 |
| $M$ | Normal | 2000 | 400 |
| $P$ | Normal | 500 | 100 |

### 7.2. Short column engineering problem

The *Short Column Function* [5,6,16] is a benchmark problem widely used in uncertainty quantification and reliability analysis. It models the behavior of a short structural column with uncertain material properties, subjected to random loads. The performance function is defined as

$$f(Y, M, P) = 1 - \frac{4M}{bh^2Y} - \frac{P^2}{b^2h^2Y^2} + 0 \cdot X_4, \tag{15}$$

where $Y$ is the yield stress, $M$ is the bending moment, $P$ is the axial force, $b$ is the width of the cross-section ($b = 5$ mm), $h$ is the depth of the cross-section ($h = 15$ mm) and $X_4$ is the fourth, redundant sample dimension. Table 1 lists the input random variables and their distributions with a correlation coefficient of 0.5 between $M$ and $P$.

This problem combines nonlinearities, correlated random variables, and material uncertainty, making it an established benchmark for evaluating methods such as polynomial chaos expansions, stochastic collocation, and reliability-based optimization.

### 7.3. Cantilever beam engineering problem

The *Cantilever Beam Functions* [5,6,40], often used in reliability analysis and uncertainty quantification, model a uniform cantilever beam subjected to horizontal and vertical loads. The responses of interest are the displacement $D(x)$ at the free end of the beam and the stress $S(x)$. They are given by

$$D(x) = \frac{4L^3}{Ewt}\sqrt{\left(\frac{Y}{t^2}\right)^2 + \left(\frac{X}{w^2}\right)^2} + 0 \cdot X_5, \tag{16}$$

$$S(x) = \frac{600Y}{wt^2} + \frac{600X}{w^2t} + 0 \cdot X_5, \tag{17}$$

where $R$ is the yield stress, $E$ is Young's modulus of the beam material, $X$ is the horizontal load, $Y$ is the vertical load, $w$ is the beam width, $t$ is the beam thickness, $L = 100$ inches is the beam length, $D_0 = 2.2535$ inches is the maximum allowed displacement, and $X_5$ is the fifth, redundant sample dimension. Similarly to the short column engineering problem, the term $0 \cdot X_5$ maintains the higher-dimensional sampling structure while the physical response depends on a reduced set of variables. This reflects the realistic case of potentially inactive inputs and enables a direct evaluation of projection robustness.

**Table 2**
Input random variables and their distributions for the cantilever beam problem.

| Variable | Distribution | Mean ($\mu$) | Std. Dev. ($\sigma$) |
|---|---|---|---|
| $R$ | Normal | 40,000 | 2,000 |
| $E$ | Normal | $2.9 \times 10^7$ | $1.45 \times 10^6$ |
| $X$ | Normal | 500 | 100 |
| $Y$ | Normal | 1000 | 100 |

The random input variables follow the distributions listed in Table 2. Modified response forms are sometimes used for defining safe regions, e.g.

$$\frac{S}{R} - 1 \le 0, \quad \frac{D}{D_0} - 1 \le 0.$$

### 7.4. Gaussian process regression and evaluation procedure

For both problems in Sections 7.2 and 7.3, the transformed inputs and the corresponding response values are used to train a Gaussian Process (GP) regression model, which is equivalent to Kriging in geostatistics and engineering. The GP model employs a Matérn-5/2 kernel with automatic relevance determination to capture different sensitivities of each input parameter. A small, fixed noise variance is assumed to prevent numerical instabilities. The kernel hyperparameters are optimized by maximizing the log-marginal likelihood.

This procedure yields a probabilistic surrogate model that provides predictions for new input samples together with an associated uncertainty measure. To quantify the predictive accuracy, we compute the root mean squared error (RMSE) of the GP predictions using a common, high-fidelity testing set of input points, at which the exact response values are available.

The left and right panels in Fig. 10 show the mean RMSE of the GP predictions for the short column and cantilever beam problems, respectively. Each thick curve corresponds to one sampling strategy and is obtained by averaging the RMSE over 500 independent realizations of the design, and is accompanied by ± one standard deviation band. For each realization, the GP model is trained on the corresponding design points, and the prediction error is evaluated with respect to the universal testing set. This setting allows a fair comparison of the different design strategies in realistic engineering applications.

In addition, corresponding RMSE plots are provided in Fig. 11 for cases in which the engineering problems are solved using input point samples without redundant dimensions, i.e., $N_{\mathrm{var}} = 3$ for the first example and $N_{\mathrm{var}} = 4$ for the second example.

### 7.5. FEM engineering problem

The last presented engineering example concerns a study of mechanical response of a meso-scale finite element model of concrete. The modeling approach was presented in Mašek and Miarka [19]. The detailed model captures explicitly the individual aggregate and matrix volumes. For aggregate and matrix, damage plasticity microplane material model is used to allow for damage within the materials. Additionally, the aggregate surfaces are covered with oriented cohesive elements allowing for phase separation.

A standard three-point bending configuration is adopted. The specimen geometry is characterized by a support span of $S = 240$ mm, a depth $D = 80$ mm, and a thickness $T = 80$ mm. In the notched configuration, the notch depth is defined as $a = 0.1D$. The notch width is 3 mm and the notch tip is rounded, as illustrated in Fig. 12. The boundary conditions follow a three-point bending test setup. The support boundary conditions are represented by defining rigid regions (CERIG elements, i.e., constraint equation rigid regions in Ansys software) on the bottom surface of the beam. Each rigid region is connected to a master node (MASS21 – point mass element) positioned at its centroid, indicated in red in Fig. 12. The separation between the two master support nodes is equal to the beam span $S$. A narrow rigid region for loading is defined along the top surface of the specimen, with a master node positioned at its center. A vertical unit load of $F_y = 1$ N is applied at the master node, while the horizontal displacement components $u_x$ and $u_z$ are constrained.

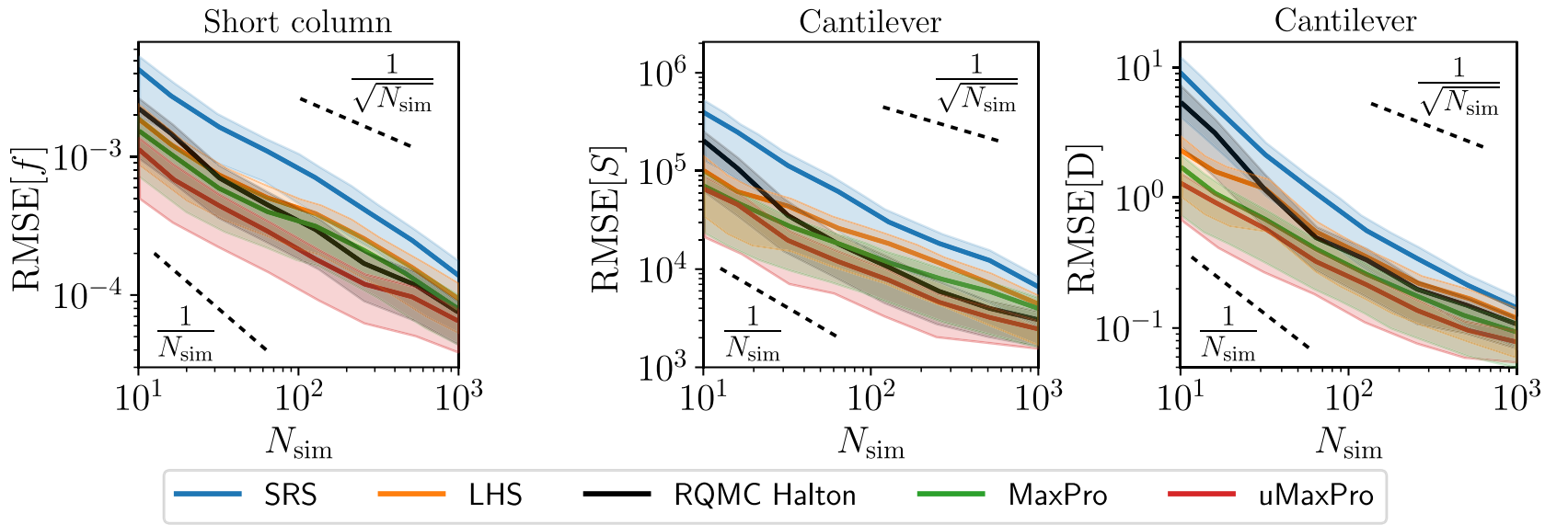

**Fig. 10.** Mean RMSE of Gaussian process predictions for the engineering examples. The left panel corresponds to the short column response. The middle and right panels correspond to the cantilever beam problem, showing the prediction errors for the stress $S$ and the displacement $D$, respectively. The shaded bands represent ± one standard deviation across 500 independent realizations.

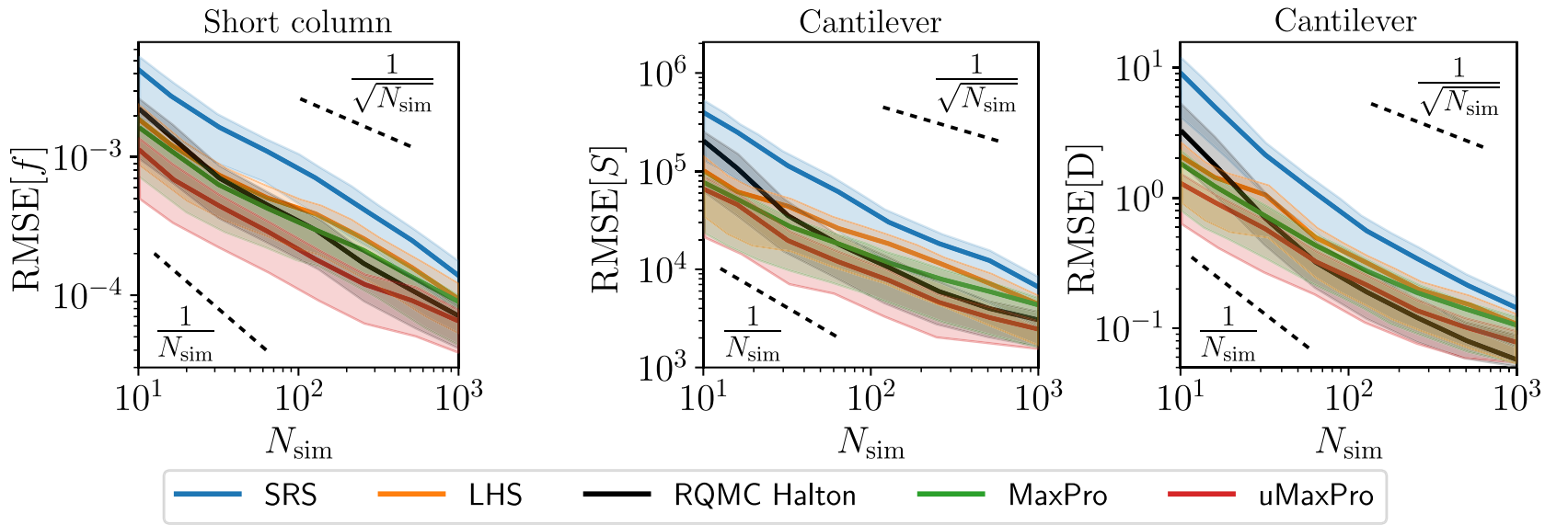

**Fig. 11.** Mean RMSE of Gaussian process predictions for the engineering examples as solved with point samples with *no redundant dimensions*. The left panel corresponds to the short column response. The middle and right panels correspond to the cantilever beam problem, showing the prediction errors for the stress $S$ and the displacement $D$, respectively. The shaded bands represent ± one standard deviation across 500 independent realizations.

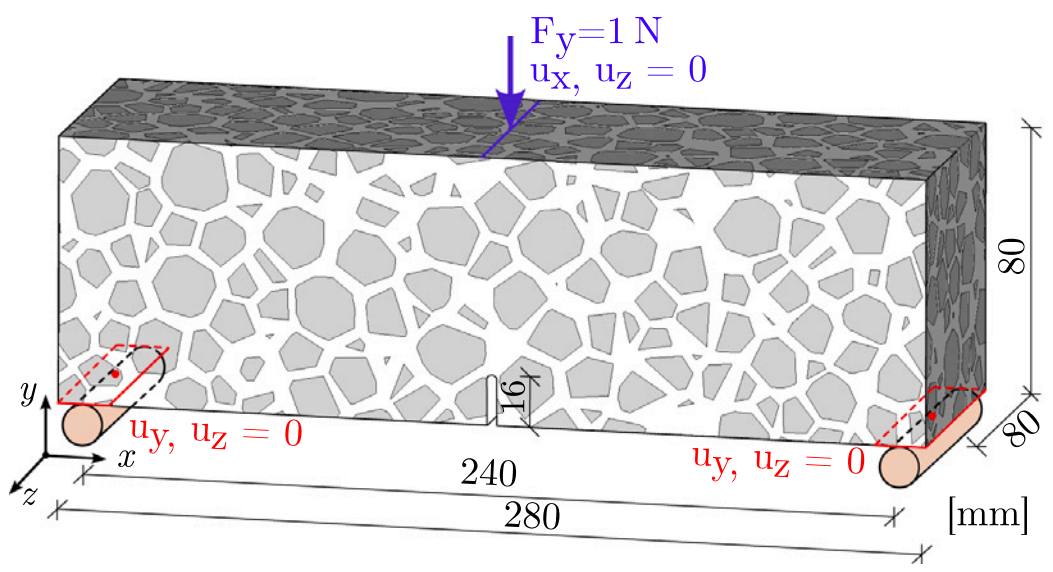

**Fig. 12.** Three-point bending specimen: boundary conditions and dimensions.

The Young's modulus of granite aggregate, cement paste, and interfacial transition zone (ITZ) was modeled using lognormal probability distributions to account for inherent material heterogeneity and positive physical constraints, see Table 3 for the input parameter distributions. The selected parameter ranges are based on experimental measurements reported in Refs. [12,23,30,39]. Due to the common hydration origin and microstructural dependency between the cement matrix and the ITZ, their elastic moduli are statistically correlated. Previous nanoindentation studies indicate that the ITZ modulus typically ranges between 40–70% of the matrix modulus and exhibits a strong positive correlation with matrix stiffness [23,39]. Therefore, a dependent random variable formulation was adopted to ensure physically consistent material representation.

The aim of this engineering example is to study the interaction of the aggregate, matrix and ITZ Young's moduli and its influence on the stress distribution within the material. The presented example attempts to estimate the 95th percentile of maximum tensile stress in the three mesostructure phases, i.e. within aggregate, matrix and ITZ materials: first principal stresses $\sigma^{95\text{th}}_{1,\text{Agg}}$ and $\sigma^{95\text{th}}_{1,\text{Mx}}$ and the maximum normal traction $\sigma^{95\text{th}}_{\text{SX,ITZ}}$.

To highlight the subspace projection qualities, we utilize randomly selected 3D subspaces of 5D point samples of the each sampling method. Fig. 13 shows the decrease of root mean square error in the estimation of $\sigma^{95\text{th}}_{1,\text{Agg}}$, $\sigma^{95\text{th}}_{1,\text{Mx}}$ and $\sigma^{95\text{th}}_{\text{SX,ITZ}}$, respectively. Clearly, the benefit of using uMaxPro over MaxPro samples is evident. However, being limited to the LHS structure, the slope of RMSE decrease for uMaxPro cannot be improved beyond MaxPro, LHS, or even SRS samples. Nevertheless, while the RQMC Halton samples inherently exhibit a steeper RMSE decrease, their low uniformity in small samples is a strong penalty that vanishes for rather high sample sizes. In engineering applications, computational cost is typically minimized, and therefore the sample size is kept as small as possible. Consequently, the uMaxPro samples offer a clear advantage in terms of the estimation precision-to–computing time ratio.

**Table 3**
Statistical parameters and correlation assumptions for Young's modulus of concrete phases.

| Material Phase | Distribution | Mean (GPa) | COV (%) | Correlation Coefficient ($\rho$) |
|---|---|---|---|---|
| Granite Aggregate | Lognormal | 60.0 | 17.5 | – |
| Cement Matrix (Paste) | Lognormal | 27.5 | 22.5 | 0.70 with ITZ |
| Interfacial Transition Zone (ITZ) | Lognormal | 17.5 | 30.0 | 0.70 with Matrix |

## 8. Conclusion

The MaxPro criterion was originally proposed as a sample optimization method that maximizes space-filling properties not only in the full design domain but also in all subsets of factors. This characteristic is crucial when only a subset of the factors is active—often the case in practice—especially when the active factors are unknown a priori. In applications such as numerical integration or function approximation, it is highly desirable to both reduce the estimation variance and ensure unbiased results.

In the present work, we have demonstrated that the original MaxPro criterion does not yield statistically uniform samples even when combined with Latin hypercube sampling. It systematically oversamples domain boundaries and undersamples domain corners. As a consequence, such nonuniformity leads to biased estimations and slow convergence in numerical integration. Moreover, when these nonuniform samples are projected onto lower-dimensional subspaces, the imbalance is exacerbated, further degrading the sample quality.

To address these issues, we introduced the periodic variant—referred to as uMaxPro—which employs a periodic design domain based on the minimum image convention. This modification ensures that point samples visit all regions of the domain with equal probability, thereby achieving both statistical uniformity and optimized space-filling properties. As a result, uMaxPro facilitates unbiased numerical integration and screening with reduced variance. Indeed, our experiments show that in subspaces, uMaxPro samples perform exceptionally well, even outperforming the Halton sequence.

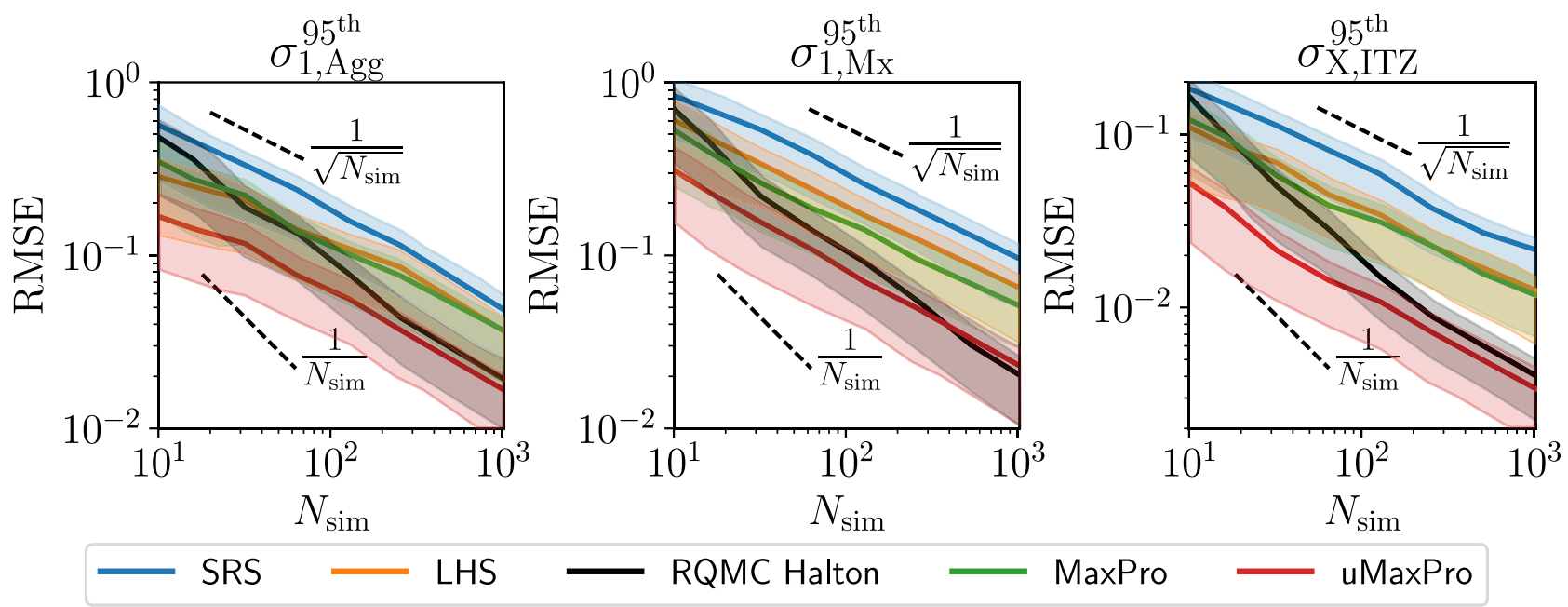

**Fig. 13.** Root mean square error of estimation of 95th quantiles of maximum tensile stress attained within aggregate, matrix and ITZ phases, respectively. The shaded bands represent ± one standard deviation.

Beyond synthetic test functions, we further demonstrated the effectiveness of uMaxPro on two benchmark engineering problems drawn from structural mechanics: the ultimate load capacity of a short column and the displacement of a cantilever beam. In both cases, the test functions depend on one dimension fewer than the optimized design, making them particularly challenging for distance-based sampling methods. The results confirmed that uMaxPro maintains its superior projection properties in these realistic engineering contexts, where standard designs provide no control over lower-dimensional subspaces.

In summary, the periodic modification embodied in uMaxPro represents a crucial improvement over the original MaxPro criterion. It not only fulfills its intended goal of strong performance in function integration and variance reduction, but also proves effective in challenging engineering applications where active subspaces of reduced dimensions are of primary importance.

## CRediT authorship contribution statement

**Miroslav Vořechovský:** Writing – review & editing, Visualization, Supervision, Software, Methodology, Conceptualization. **Jan Mašek:** Writing – original draft, Visualization, Methodology, Investigation, Data curation, Conceptualization.

## Declaration of competing interest

The authors declare that they have no known competing financial interests or personal relationships that could have appeared to influence the work reported in this paper.

## Acknowledgments

The authors acknowledge the financial support provided by the Czech Science Foundation under project # GF22-06684K.

The development of the software for constructing optimized designs was funded by the European Union via the OP JAK project INODIN (CZ.02.01.01/00/23_020/0008487).

Numerical calculations were performed within the IT4Innovations National Supercomputing Center, Czech Republic supported by the Ministry of Education, Youth and Sports of the Czech Republic through the e-INFRA CZ (ID:90254).

The authors also thank Martin Vořechovský for his assistance in implementing the software used to generate the proposed designs.

## Data availability

Research Link Provided